\documentclass[sn-basic,default,iicol]{sn-jnl}

\usepackage{graphicx}%
\usepackage{multirow}%
\usepackage{amsmath,amssymb,amsfonts}%
\usepackage{amsthm}%
\usepackage{mathrsfs}%
\usepackage[title]{appendix}%
\usepackage{xcolor}%
\usepackage{textcomp}%
\usepackage{manyfoot}%
\usepackage{booktabs}%
\usepackage{algorithm}%
\usepackage{algorithmicx}%
\usepackage{algpseudocode}%
\usepackage{listings}%
\usepackage{geometry}
\usepackage{makecell}
\usepackage{tabu}
\usepackage{graphicx}
\usepackage[T1]{fontenc}
\usepackage[utf8]{inputenc}

\usepackage{mathptmx}
\usepackage{setspace}
\usepackage{stfloats}
\usepackage{diagbox}

\usepackage[section]{placeins}

\usepackage{picins}

\theoremstyle{thmstyleone}%
\theoremstyle{thmstyletwo}%

\theoremstyle{thmstylethree}%

\begin{document}

\title[Article Title]{Multi-service collaboration and composition of cloud manufacturing customized production based on problem decomposition}


\author[1]{\fnm{Hao} \sur{Yue}}\email{hyue@upc.edu.cn}

\author*[1]{\fnm{Yingtao} \sur{Wu}}\email{z22070143@s.upc.edu.cn}

\author[2]{\fnm{Min} \sur{Wang}}\email{wangmin1@yzu.edu.cn}
\author[3,4]{\fnm{Hesuan} \sur{Hu}}\email{huhesuan@gmail.com}
\author[5,6]{\fnm{Weimin} \sur{Wu}}\email{wmwu@iipc.zju.edu.cn}
\author[7,8]{\fnm{Jihui} \sur{Zhang}}\email{zhangjihui@qdu.edu.cn}


\affil*[1]{\orgdiv{College of Computer Science and Technology}, \orgname{China University of Petroleum (East China)}, \orgaddress{\city{Qingdao}, \postcode{266580}, \state{Shandong}, \country{China}}}
\affil[2]{\orgdiv{College of Information Engineering}, \orgname{Yangzhou University},\orgaddress{\city{Yangzhou}, \postcode{225127}, \state{Jiangsu}, \country{China}}}
\affil[3]{\orgdiv{School of Electro-Mechanical Engineering}, \orgname{Xidian University},\orgaddress{\city{ Xi’an}, \postcode{710071}, \state{Shanxi}, \country{China}}}
\affil[4]{\orgdiv{School of Computer Science and Engineering, College of Engineering}, \orgname{Nanyang Technological University},\orgaddress{\postcode{ 639798}, \country{Singapore}}}
\affil[5]{\orgdiv{State Key Laboratory of Industrial Control
Technology}, \orgname{Zhejiang University}, \orgaddress{ \city{Hangzhou}, \postcode{310027}, \state{Zhejiang}, \country{China}}}
\affil[6]{\orgdiv{Institute of Cyber-Systems and Control}, \orgname{Zhejiang University}, \orgaddress{ \city{Hangzhou}, \postcode{310027}, \state{Zhejiang}, \country{China}}}
\affil[7]{\orgdiv{Institute of Complexity Science, School of Automation}, \orgname{Qingdao University}, \orgaddress{ \city{Qingdao}, \postcode{266071}, \state{Shandong}, \country{China}}}
\affil[8]{\orgname{Shandong Key Laboratory of Industrial Control Technology}, \orgaddress{ \city{Qingdao}, \postcode{266071}, \state{Shandong}, \country{China}}}

\abstract{
Cloud manufacturing system is a service-oriented and knowledge-based one, which can provide solutions for the large-scale customized production. The service resource allocation is the primary factor that restricts the production time and cost in the cloud manufacturing customized production (CMCP). In order to improve the efficiency and reduce the cost in CMCP, we propose a new framework which considers the collaboration among services with the same functionality. A mathematical evaluation formulation for the service composition and service usage scheme is constructed with the following critical indexes: completion time, cost, and number of selected services. Subsequently, a problem decomposition based genetic algorithm is designed to obtain the optimal service compositions with service usage schemes. A smart clothing customization case is illustrated so as to show the effectiveness and efficiency of the method proposed in this paper. Finally, the results of simulation experiments and comparisons show that these solutions obtained by our method are with the minimum time, a lower cost, and the fewer selected services.}



\maketitle

\section{Introduction}\label{sec1}

Cloud manufacturing (CMfg) refers to a service-oriented networked manufacturing model that manages decentralized and virtualized manufacturing resources, capabilities, and knowledge to provide on-demand manufacturing cloud services to customers \cite{2022WangWenbo(1),2022WangMin(2)}. With the widespread application of cutting-edge technologies like industrial internet of things (IoT), cyber-physical systems, and cloud computing, the traditional manufacturing model is transforming into a smart manufacturing model\cite{2022ZhangYingfeng(3)}. As an emerging manufacturing paradigm, CMfg is able to handle more complex manufacturing requests to satisfy multi-objective demands than the traditional manufacturing modes. CMfg is progressively becoming an important means of upgrading and transforming the manufacturing industry. On the CMfg service platform, customers' demands evolve from simple needs to personalized requirements with multi-dimensional aspects such as product configuration, attributes, quality, service features, timelines, and so on. Distributed resource integration and integrated resource distribution are all reflected in CMfg. The CMfg service model can integrate distributed manufacturing resources and overcome traditional manufacturing limitations of providing only a single service. In addition, the CMfg service model can redistribute integrated manufacturing resources and improve the efficiency of resource utilization. CMfg services combine advanced computing, intelligent optimization, virtualization, embeddedness, the IoT, and other high-performance technologies to optimize industrial structures and rationalize resource utilization\cite{2022PanQK(44),2016QuT(6)}. For example, virtualization technologies construct virtual manufacturing environments to simulate system resources and capabilities, reducing wastage during manufacturing \cite{2017ZhongRayY(4)}. There are many production management strategies that can improve the efficiency of CMfg, such as capturing resource states, optimizing production decisions through intelligent algorithms, delivering timely and accurate instructions \cite{2022HuHao(5)}, and enabling real-time scheduling \cite{2016QuT(6)} and logistics synchronization.

Cloud manufacturing customized production (CMCP) is the concept of providing products and services tailored to customers' individual requirements \cite{2022WangMin(2),2019LuYuqian(9)}. In the traditional manufacturing model, customized production often faces challenges such as uncontrollable production costs, unpredictable delivery dates, and unguaranteed product quality. However, CMfg has brought a turnaround to the development of customized manufacturing. Customized production in CMfg represents an intelligent manufacturing model where customers participate in the entire service lifecycle, including product customization, design, development, manufacturing, and logistics \cite{2019LuYuqian(9)}. The CMfg platform can serve as a center for a product development, resource scheduling and management, product trading, and progress monitoring. It provides a strong guarantee for the implementation of customized manufacturing.

For example, China Changan Automobile Group has built an online platform using bill of materials management and product data management to enable personalized customization of automobiles \cite{changan}. By integrating dozens of manufacturing bases and factories worldwide, the production cycle is effectively shortened and the indirect inventory is reduced. For personalized vehicle customization, the shortest lead time from customer order to receipt of the complete vehicle is reduced to 7 days, and over 15,000 personalized configurations are made available to customers. MindSphere is a cloud-based IoT operating system from Siemens that helps customers to source, transport, store, analyze, and manufacture \cite{MindSphere}. The platform provides an open application programming interface for selecting applications that analyze factory data and intelligently control production. Established by Haier, COSMOPlat is an industrial internet platform with four architectural layers: resources, platform, applications, and models \cite{COSMOPlat}. The platform layer consists of seven modules: user interaction customization, marketing, open design, purchasing, intelligent production, intelligent logistics, and intelligent service. Users can participate in design, procurement, manufacturing, and logistics. 
They can directly experience the production process. It creates a trinity of users, enterprises, and resources. These examples show architectural designs for CMfg customized services. 

Service composition optimal selection (SCOS) is the selection of a series of cloud services from candidate service sets to accomplish complex tasks according to certain processes and rules\cite{2022GaoYifan(13)}. That is the paramount issue for CMfg. In the face of diversified user demands, the premise of SCOS is to accurately analyze different situations and problems in multitask requests \cite{2020YuanM(8)}. The key to service composition is to select the best service to meet user requirements from massive and aggregated candidate services \cite{2019LuYuqian(9)}. The cloud service composition can be described as four basic structures: sequential, parallel, selective, and circular. A sequential structure refers to a unidirectional sequential dependency among sub-tasks, where each sub-task starts after the completion of the preceding sub-task. Parallel, selective, and circular structures can be converted into sequential structures by mature modern technologies \cite{2017ZhouJiajun(10)}.

Compared with ordinary production in CMfg, customized production focuses not only on the manufacturing process but also on the customized design of the product. The output of the customized design phase is usually digital rather than physical. These digital products can be effectively managed by a cloud platform and delivered to subsequent services in real time. It opens up more possibilities for the service composition of customized production. For example, in the sequential structure, multiple cloud services with identical functionality collaborate to perform a sub-task, where the next sub-task starts after the previous one starts, not after the previous one finishes. Meanwhile, events such as cloud service outages or the release of busy cloud services are possible in real-world production scenarios.
All of this complicates an already complex situation of the SCOS problem for customized production. Although the number of services in the cloud service pool is very large, it is still difficult to satisfy customized production tasks well.

Based on the virtualization model of cloud service, some traditional intelligent optimization algorithms, such as ant colony optimization (ACO) \cite{2017ZhouJiajun(23)}, gray wolf optimizer (GWO) \cite{2019YangYefeng(22)}, particle swarm optimization (PSO) \cite{2019Zhang(25)}, and genetic algorithm (GA) \cite{2017Karimi(24)}, are widely used and well studied for solving the SCOS problem \cite{2020ding(11)}. However, almost all of these traditional intelligent algorithms have their respective limitations. Specifically, ACO is more commonly used for path optimization, GWO has high convergence and search efficiency, but poor stability, PSO is more suitable for continuous variable optimization, and for GA, the solution space becomes very large when the dimensionality of the independent variables is high. In CMfg, the SCOS usually has a large and irregular solution space when the production is complex and on a large-scale. This makes it difficult to find the optimal solution, and causes the conventional intelligent optimization algorithms to converge inefficiently and easily fall into a local optimum. The SCOS for customized production is more complex than its counterpart for an ordinary case. There is a lack of research on customized production SCOS in CMfg. As a consequence, it is imperative to design a highly efficient intelligent algorithm to solve the customized production SCOS problem.

Considering the above problems and situations, this paper focuses on CMfg services for customized production, especially together with the collaboration of multiple services on the same sub-task. A framework for customized production on the
CMfg system is proposed. 
We develop an improved GA for the customized production. Experiments show the effectiveness and efficiency of the our method.
Specially, the main contributions are as follows.

1) A framework for the CMCP is proposed in order to allow multiple services to collaborate on the same sub-task.

2) A mathematical evaluation formulation is constructed in the proposed framework to search for the optimal service composition and service usage scheme.

3) A problem decomposition based GA is designed to obtain the optimal service composition and service usage scheme.

The remaining sections are organized as follows: Section\:\ref{sec2} reviews related work. Section\:\ref{sec3} introduces the CMfg service system, the proposed customized production framework, and the optimization goal. The problem decomposition based GA is presented in Section\:\ref{sec4}. Section\:\ref{sec5} gives experimental simulation results. Section\:\ref{sec6} compares the difference between our method and some existing studies. Finally, Section\:\ref{sec7} concludes this paper.

\section{Related work}\label{sec2}

In recent years, much effort has been put into developing and implementing methods for solving CMfg SCOS. Many improved intelligent algorithms are applied to large-scale SCOS problems due to their effectiveness, simplicity, and generality \cite{2022Thakur(12)}.

By combining multiple service quality metrics into one through specified weights, most of the earlier research focused on single-objective optimization.
For example, Li et al. \cite{2020Li(18)} proposed an extended Gale-Shapley algorithm-based approach that can effectively generate multiple quality of service (QoS)-aware service compositions.  
Ding et al. \cite{2020ding(11)} designed a niching behavior-based gravitational search algorithm to address manufacturing SCOS problem. 
Additionally, some hybrid algorithms were proposed.
Bouzary and Chen \cite{2019Bouzary(26)} proposed a new hybrid approach based on the developed gray wolf optimizer algorithm and evolutionary operators of the GA.
Liu et al. \cite{2021LiuZ(16)} proposed a novel hybrid algorithm to address the personalized recommendation for manufacturing service composition.  

Recently, several researchers have studied optimization methods for CMfg SCOS from a multi-objective perspective \cite{2019Liu(46)}. These methods are not limited to QoS or ordinary SCOS.  
In \cite{2019Zhang(25)}, a PSO-based multi-objective algorithm was proposed to include more feasible service composition candidates. 
The nondominated sorting genetic algorithm II (NSGA-II) \cite{2002Kalyanmoy(30)} is a promising multi-objective algorithm, which was used in the reference \cite{2022WangMin(2)} because of its good global exploration performance and strong stability. However, the NSGA-II for large-scale problems may get trapped in a local optimum.
Seghir \cite{2021Seghir(15)} proposed a fuzzy discrete multi-objective artificial bee colony approach to solving the multi-objective QoS-driven web service composition problem. 
Yaghoubi and Maroosi \cite{2020Yaghoubi(17)} proposed an improved multi-verse optimization algorithm for web service composition to improve QoS while satisfying the service level agreement. 

For the purpose of energy conservation, an energy-aware service composition mechanism was proposed in \cite{2020Ibrahim(20)} to minimize the energy consumption of mobile cloud providers and a hybrid shuffled frog leaping algorithm and genetic algorithm were used to optimize mobile cloud service composition. In addition, Yang et al. \cite{2020YangYefeng(19)} presented an enhanced multi-objective gray wolf optimizer for the multi-objective SCOS in CMfg, where both the QoS and the energy consumption are considered from the perspective of sustainable manufacturing. 

To efficiently solve the robust SCOS problem, a guiding artificial bee colony-gray wolf optimization algorithm with three improvement strategies was proposed in \cite{2020Yang(21)}. 
Gao et al. \cite{2022GaoYifan(13)} developed a strengthened multi-objective gray wolf optimizer to solve the bi-objective SCOS problem which considers both robustness and QoS. 

For the SCOS in uncertain environments, Zhang et al. \cite{2021ZhangS(14)} proposed a new genetic based hyper-heuristic algorithm with adjustable chromosome length for the multitask-oriented manufacturing service composition model. The hybrid algorithm consists of a clustering-based collaborative filtering algorithm and an improved personalization-oriented third generation non-dominated sorting genetic algorithm. 

The above research results from the existing literature have made great progress in solving the SCOS problem. Mass personalization are also feasible \cite{2018Xiong(32),2021Huo(34),2022TaoFei(43)}. Kang et al. \cite{2023Kang(33)} investigated the 3D printing service allocation problem with optimization-based and real-time allocation strategies. 
Aheleroff et al. \cite{2021Aheleroff(35)} addressed mass personalization as part of manufacturing to meet unique and complex requirements on a large scale by harnessing Industry 4.0 technologies. 
Wang et al. \cite{2022WangMin(2)} established a CMfg service model based on rewritable Petri nets to describe and analyze the reconfiguration process of the CMCP. Additionally, a CMfg resource allocation strategy with service collaboration in mind is established. However, there has been no improvement in the intelligent algorithm that is used in the strategy for the CMCP. In summary, the algorithm for the CMCP is limited and few studies consider multi-service collaboration with the same functionality in CMfg.

In view of the above discussion, this paper designs a CMCP framework with collaboration among similar services to increase productivity, and proposes an improved algorithm to obtain the service compositions and the associated service usage schemes.

\section{Proposed framework and evaluation formulation for CMCP}\label{sec3}

In this section, the CMfg service system is introduced. Then, the framework with service collaboration for CMCP is proposed. Finally, the evaluation formulation of service composition and service usage scheme is presented.

\subsection{Cloud manufacturing service system}\label{subsec3-1}



\begin{figure}[htbp]
\centering 
\includegraphics[scale=0.6]{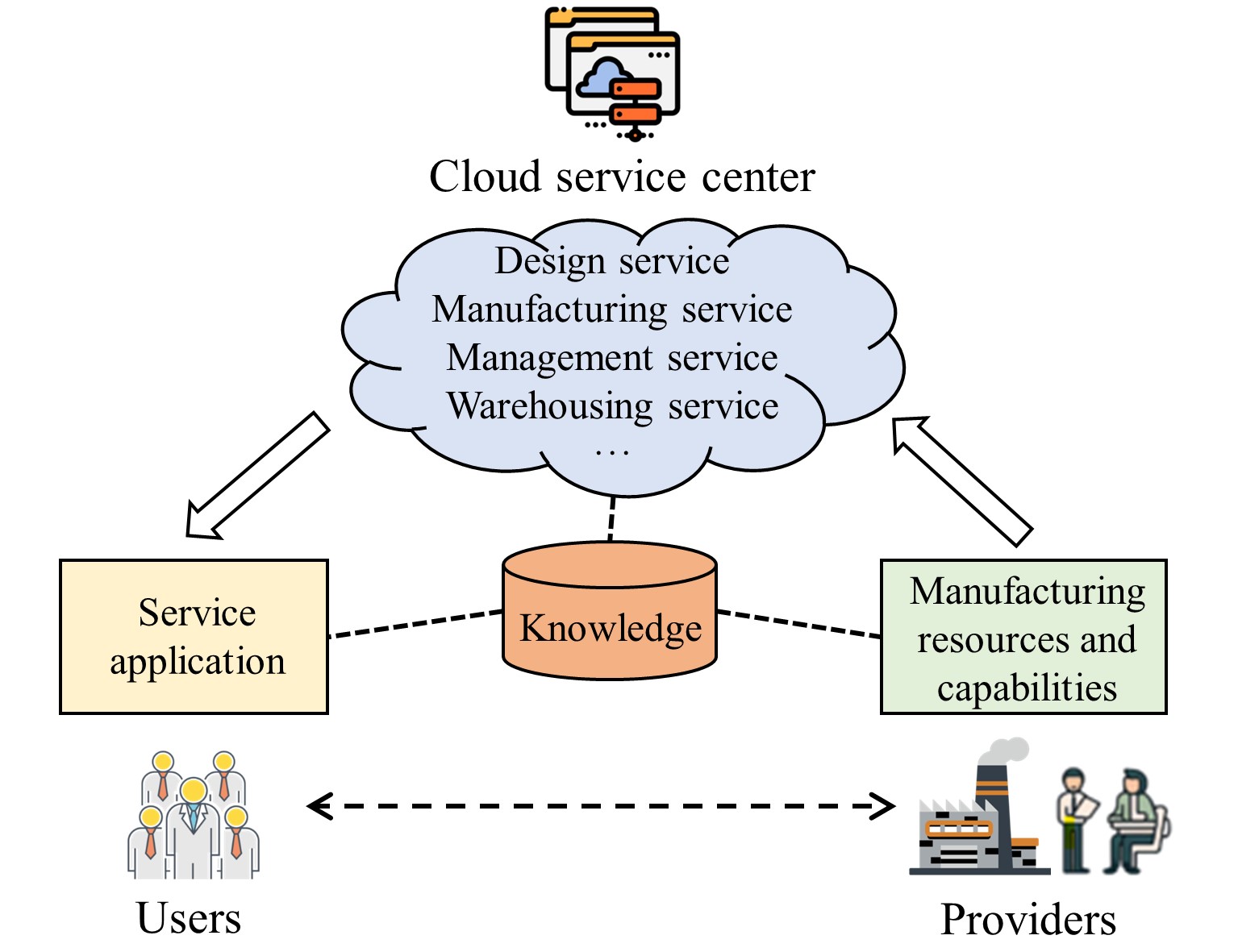}
\caption{\small CMfg service system consisting of three roles}
\label{fig:Fig.2}
\end{figure}

As shown in Fig.\:\ref{fig:Fig.2}, the CMfg service system consists of three roles: provider, cloud service center, and user. A system includes multiple providers, which provide physical manufacturing resources and capabilities to the cloud service center. Manufacturing resources include soft, hard, and human resources \cite{2014Huang(31)}. Manufacturing capabilities are expressed by combinations of manufacturing resources. Manufacturing capabilities reflect the ability to complete a manufacturing task supported by related manufacturing resources. Cloud service center encapsulates resources and capabilities of providers into cloud services, and then efficiently manages and operates cloud services. Different providers with different manufacturing resources and capabilities may offer some cloud services with identical functionality and different QoS, including time, cost, etc. 
\vspace{-0.4em}
Cloud services are delivered to users on demands. Users consume different services as needed with the support of the cloud service center.  
Knowledge enables virtual access to manufacturing resources and capabilities through service-oriented encapsulation and categorization, along with efficient management and intelligent search of cloud services. In addition, for special cases or simple manufacturing tasks, users can directly request services from providers, while the latter can directly offer services to the former. 

\subsection{The proposed framework for cloud manufacturing customized production}\label{subsec3-2}
\hspace{1em}\textbf{Definition 1} \cite{2022WangMin(2)} (Order) An order in CMfg is a four-tuple $O=(id, type,$ $quant\!\!\:it\!\!\:y)$, where $id$ represents the id of order, $type$ represents the type of products, and $quant\!\!\:it\!\!\:y$ represents the quantity of products.

\begin{figure}[htbp]
\flushright
\includegraphics[scale=0.6]{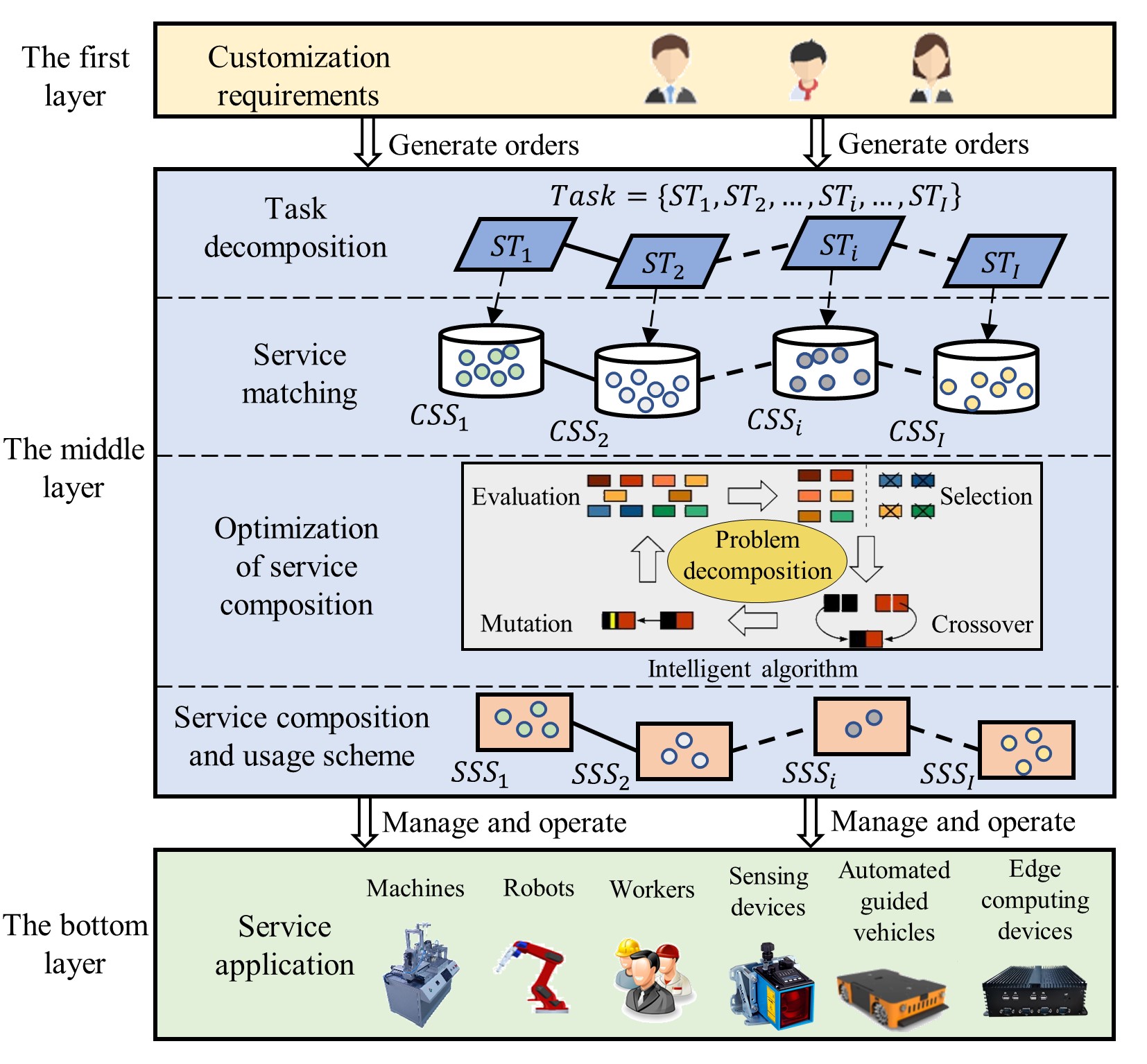}
\caption{\small Customized production framework on CMfg service system}
\label{fig:Fig.3}
\end{figure}

The design phases of customized production output digital intermediate products instead of physical products. These digital products can be efficiently managed through the cloud service center and delivered to be processed by subsequent services in real time. This opens up more possibilities for the service composition of customized production. 
As shown in Fig.\:\ref{fig:Fig.3}, the customized production framework with service collaboration in mind consists of three layers. 
The first layer is the user layer, which collects customization requirements and generates orders. An order is considered as one task. 
The middle layer is the cloud service center layer, which receives the user orders, decomposes the tasks into sub-tasks, and matches the sub-tasks with the candidate services to generate service compositions as well as service usage schemes. 
The bottom layer is the service application layer, which performs product design and manufacturing to meet the customization requirements. 
For ease of being understood, the following concepts are described: 

1) Task decomposition: A task, denoted by $Task$, can be decomposed into multiple sub-tasks $ST$. That is, $Task=\{{ST}_1,\ {ST}_2,\:\cdots,\ {ST}_{\!i},\:\cdots,\ {ST}_{\!I}\}$.

2) Service matching: Each sub-task ${ST}_{\!i}$ corresponds to a candidate service set ${CSS}_i$ containing services with identical functionality but different costs and times of a single use. That is, ${CSS}_i=\{{CS}_{i,1},\ {CS}_{i,2},\:\cdots,\ {CS}_{i,j},\:\cdots,\ {CS}_{i,\,J}\}$ where ${CS}_{i,j}$ denotes a candidate service, $j=1,\ 2,\:\cdots,\ J$.

3) Collaboration: Services can collaborate on a sub-task. The framework allows services with the same functionality to collaborate.

4) Service composition: For each sub-task ${ST}_{\!i}$ and the corresponding candidate service set ${CSS}_i$, $i=1,\ 2,\:\cdots,\ I$, there exists a selected service set ${SSS}_i=\{{CS}_{i,k},\:\cdots,\ {CS}_{i,m},\:\cdots,\ {CS}_{i,n}\}$, where ${SSS}_i$ is a non-empty subset of ${CSS}_i$. A service composition can be denoted as a set $\{{SSS}_1,\ {SSS}_2,\:\cdots,\ {SSS}_i,\:\cdots,\ {SSS}_I\}$. 

5) Service usage scheme: The services in ${SSS}_i$ for the sub-task ${ST}_{\!i}$ have the same functionality, $i=1,\ 2,\:\cdots,\ I$. The service usage scheme indicates how many times each of these services will be used.

6) Optimization: When conducting a service composition and its usage scheme, there are always some constraints such as time, cost, and limited resources. Under various constraints, in order to satisfy the requirements as much as possible, the service composition and service usage scheme should be optimal. 

Compared with general frameworks for cloud service composition such as those presented in \cite{2017ZhouJiajun(23)} and \cite{2019Bouzary(26)}, the framework proposed in this paper allows multiple services with the same functionality to collaborate on the same sub-task. As a consequence, our proposed framework lays a solid foundation for improving productivity and reducing resource idleness.

\vspace{-1.5em}

\subsection{Optimal service composition and service usage scheme}\label{subsec3-3}

The optimal CMCP service composition and the associated service usage scheme should be of high efficiency and low consumption, and use a minimal amount of service resources. Therefore, the optimization goal consists of three objectives: the minimum total completion time, the lowest total production cost, and the minimum number of selected services. The total completion time (resp., total production cost) refers to the total time (resp., cost) required to complete a task. The number of selected services indicates how many services are selected to complete a task.

For example, there is a simple custom order with the $quant\!\!\:it\!\!\:y$ equal to 10. The order is treated as a production task, which contains two sub-tasks: design sub-task ${ST}_d$ and manufacturing sub-task ${ST}_m$. There are two candidate services ${CS}_{d,1}$ and ${CS}_{d,2}$ for ${ST}_d$, and two candidate services ${CS}_{m,1}$ and ${CS}_{m,2}$ for ${ST}_m$. Assume that the four services are with the same time of a single use. The service ${CS}_{d,1}$ (resp., ${CS}_{m,1}$) has a lower cost per use than the service ${CS}_{d,2}$ (resp., ${CS}_{m,2}$). There are the following nine service compositions for this task: 
$\{{CS}_{d,1},\ {CS}_{m,1}\}$, 
$\{{CS}_{d,1},\ {CS}_{m,2}\}$, 
$\{{CS}_{d,2},\ {CS}_{m,1}\}$, 
$\{{CS}_{d,2},\ {CS}_{m,2}\}$, 
$\{{CS}_{d,1},\ {CS}_{m,1},\ {CS}_{m,2}\}$, 
$\{{CS}_{d,2},\ {CS}_{m,1},\ {CS}_{m,2}\}$, 
$\{{CS}_{d,1},\ {CS}_{d,2},\ {CS}_{m,1}\}$, 
$\{{CS}_{d,1},$ $\ {CS}_{d,2},\ {CS}_{m,2}\}$, 
and $\{{CS}_{d,1},\ {CS}_{d,2},\ {CS}_{m,1},\ {CS}_{m,2}\}$. Each service composition has one or more usage schemes. A complete customized production solution consists of a service composition and a service usage schemes. To minimize the total production cost and the number of selected services, the service composition $\{{CS}_{d,1},\ {CS}_{m,1}\}$ would be selected, and services ${CS}_{d,1}$ and ${CS}_{m,1}$ should be used 10 times each. To minimize the total completion time of the task, the service composition $\{{CS}_{d,1},\ {CS}_{d,2},\ {CS}_{m,1},\ {CS}_{m,2}\}$ should be selected, and each service would be used 5 times.

For a large-scale task, it is difficult to compute the total completion time accurately. In the sequel, a fast recursive approach is proposed to approximate the total completion time. Meanwhile, the formulas for calculating the total production cost and the number of selected services are also given. We assume that several services with the same functionality are selected for sub-task ${ST}_{\!i}$, and the set ${SSS}_i$ contains selected services for ${ST}_{\!i}$, $i=1,\ 2,\:\cdots,\ I$. 
The theoretical minimum completion time of ${ST}_{\!i}$ is denoted by ${Time}_i$. The cumulative usage time of the service ${CS}_{i,j}$ in ${SSS}_i$, $j\in \{1,\ 2,\:\cdots,\ J\}$, is denoted as follows:
\begin{flalign}
&\textrm{Time}({CS}_{i,j})=\sum_{k=1}^{K}{time}_k ,&
\label{equation:Equation 1}
\end{flalign}
where $K$ is the usage number of ${CS}_{i,j}$ and ${time}_k$ is the $k$th usage time of ${CS}_{i,j}$, $k= 1,\ 2,\:\cdots,\ K$.

For each $i\in\{1,\ 2,\:\cdots,\ I\}$, let ${LCS}_i$ represent the selected service with the longest cumulative usage time in ${SSS}_i$. The maximum time of a single use for service ${LCS}_i$ is denoted by ${UT}_i$. 
The cumulative usage time of ${LCS}_i$ is denoted as:
\begin{flalign}
&\ {LT}_i=\textrm{Time}({LCS}_{i}).&
\label{equation:Equation 2}
\end{flalign}

The sub-tasks are executed sequentially. The theoretical minimum completion time ${Time}_{i}$ of ${ST}_{\!i}$ is determined by ${Time}_{i-1}$, ${UT}_{i-1}$, ${LT}_{i}$ and ${UT}_{i}$, as follows:
\begin{flalign}
\begin{split}
{Time}_i=
&\begin{cases}
{LT}_1, \!\!\!\!\!\!&\textrm{if}\ i=1,\!\!\!\!\!\!\!\!\\
\max({LT}_{i},\ {Time}_{i-1}-{UT}_{i-1}+{UT}_{i}), \!\!\!\!\!\!\!&\textrm{if}\ i\neq 1,\!\!\!\!\!\!\!\!\\
\end{cases}\\
\end{split}&
\label{equation:Equation 3}
\end{flalign}
where ${LT}_{i}$ is calculated using Eq.\:(\ref{equation:Equation 2}), $i=1,\ 2,\:\cdots,\ I$.

The cost of completing sub-task ${ST}_{\!i}$ is :
\begin{flalign}
&\ \textrm{Cost}\left({ST}_{\!i}\right)=\sum_{{CS}_{i,j}\in {SSS}_i}{\textrm{cost}({CS}_{i,j})},&
\label{equation:Equation 4}
\end{flalign}
where cost$({CS}_{i,j})\ $represents total cost of using the service ${CS}_{i,j}$.

The number of selected services in ${SSS}_i$ is:
\begin{flalign}
&\ \textrm{Num}\left({ST}_{\!i}\right)=\textrm{num}({SSS}_i),&
\label{equation:Equation 5}
\end{flalign}
where num$({SSS}_i)$ represents the number of selected services in the set ${SSS}_i$.

Recall that $Task=\{{ST}_1,\ {ST}_2,\:\cdots,\ {ST}_{\!i},\:\cdots,\ {ST}_{\!I}\}$. 
For $Task$, let ${Time}_{total}$ represent its total completion time, ${Cost}_{total}$ be the total production cost, and ${Num}_{total}$ indicate the number of services selected for $Task$. We have:
\begin{flalign}
&\ \textrm{Minimize}\ {Time}_{total}={Time}_I+\sum_{i=1}^{I-1}{UT}_i,&
\label{equation:Equation 6}
\end{flalign}
\vspace{-2em}
\begin{flalign}
&\ \textrm{Minimize}\ {Cost}_{total}=\sum_{i=1}^{I}{\textrm{Cost}\left({ST}_{\!i}\right)},&
\label{equation:Equation 7}
\end{flalign}
\vspace{-2em}
\begin{flalign}
&\ \textrm{Minimize}\ {Num}_{total}=\sum_{i=1}^{I}{\textrm{Num}\left({ST}_{\!i}\right)},&
\label{equation:Equation 8}
\end{flalign} 
where\ ${Time}_I$, Cost$\left({ST}_{\!i}\right)$, and Num$({ST}_{\!i})$ are calculated using Eqs.\:(\ref{equation:Equation 3}-\ref{equation:Equation 5}), respectively. 

\section{Intelligent algorithm for SCOS in CMCP}\label{sec4}

In this section, a problem decomposition based genetic algorithm (PDGA) is proposed to obtain the optimal CMCP service compositions and the associated service usage schemes. 

\subsection{Problem decomposition and encoding}\label{subsec4-1}

\begin{figure}[htbp]
\centering 
\includegraphics[scale=0.65]{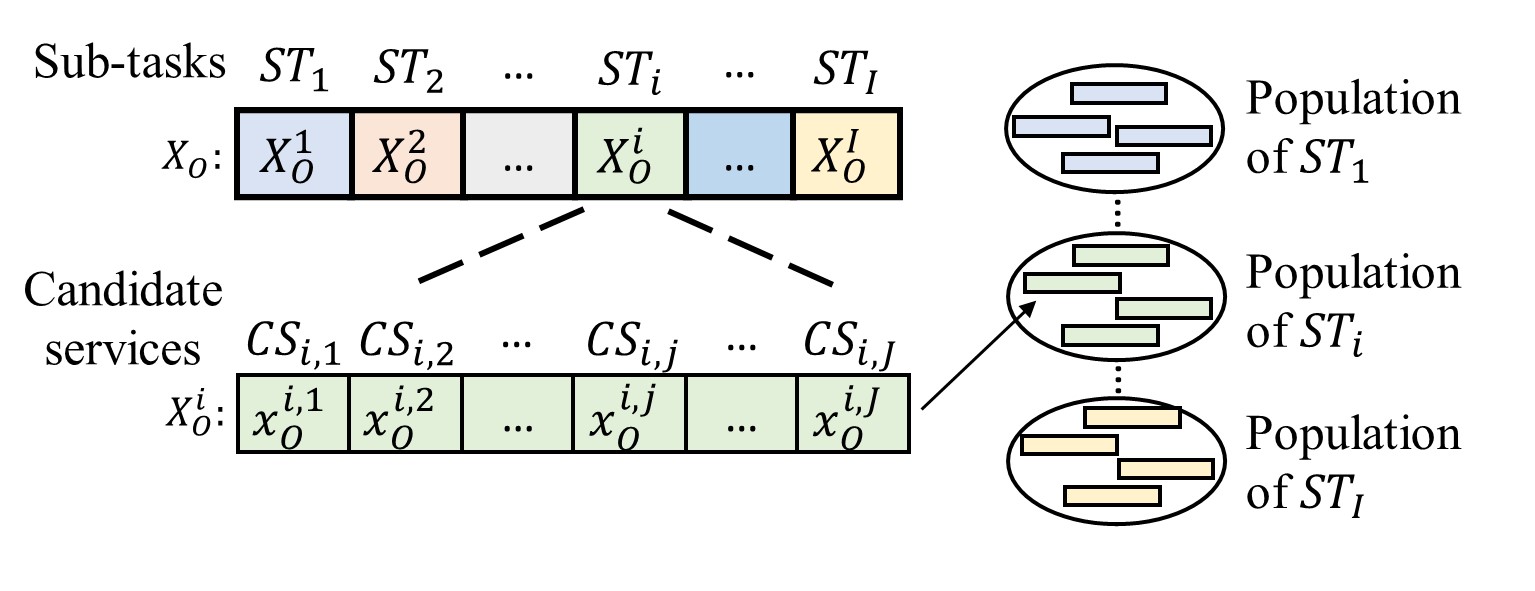}
\caption{\small Encoding for a task}
\label{fig:Fig.4}
\end{figure}

The problem decomposition can reduce the difficulty of solving complex problems by decomposing them into multiple parts. More precisely, a problem is decomposed into several parts and each part is optimized separately.  
Let $O$ represent an order. We assume that the production task of $O$ contains $I$ sub-tasks $\{{ST}_1,\ {ST}_2,\:\cdots,\ {ST}_{\!i},\:\cdots,\ {ST}_{\!I}\}$ with ${ST}_{\!i}$ being the $i$th sub-task. As shown in Fig.\:\ref{fig:Fig.4}, a service composition and its usage scheme for $O$ can be encoded as a chromosome set $X_O$ with its size equal to the number of sub-tasks. The set $X_O=\{X_O^1,\ X_O^2,\:\cdots,\ X_O^i,\:\cdots,\ X_O^I\}$ is a complete solution, and any chromosome in $X_O$ is a partial solution corresponding to a sub-task. For each sub-task, a population will be generated to find the optimal partial solution.
Thus, the complex problem of finding the optimal service composition and usage scheme for a task is decomposed into sub-problems of finding service composition and usage scheme for each sub-task.

For each $i\in\{1,\ 2,\:\cdots,\ I\}$, a chromosome $X_O^i$ with respect to sub-task ${ST}_{\!i}$ for order $O$ satisfies the following conditions:

1) $X_O^i$ is a non-zero vector with non-negative integer components. Assume that there are $J$ candidate services for the sub-task ${ST}_{\!i}$. Then, the dimension of $X_O^i$ is $J$, that is, $X_O^i= [x_O^{i,1},\ x_O^{i,2},\:\cdots,\ x_O^{i,j},\:\cdots,\ x_O^{i,\,J}]$.

2) The component $x_O^{i,j}\ge 0$ in $X_O^i$ corresponds to the $j$th candidate service ${CS}_{i,j}$, $j=1,\ 2,\:\cdots,\ J$. If $x_O^{i,j}$ is equal to zero, then ${CS}_{i,j}$ will not be selected; otherwise, ${CS}_{i,j}$ will be selected and used. Furthermore, the usage number of ${CS}_{i,j}$ is equal to $x_O^{i,j}$.

3) $\sum_{j=1}^{J}x_O^{i,j}=quant\!\!\:it\!\!\:y$, where 
$quant\!\!\:it\!\!\:y$ indicates how many products are in the order $O$.

4) Additional restrictions. For example, if a candidate service can only be used a maximum of 20 times, then the corresponding component cannot be greater than 20. 

\subsection{Algorithm overview}\label{subsec4-2}

The basic process of the problem decomposition based genetic algorithm (PDGA) is shown in Algorithm \:\hyperref[algo1]{1}.
First, the service composition and associated service usage scheme can be encoded as a set of chromosomes by the problem decomposition. For each chromosome, create a population to search for the necessary individuals. Then, combine individuals from multiple populations to obtain the solutions that are the optimal service compositions and associated service usage schemes. 
In addition, the elite selection, the fast non-dominated sorting method, and the crowding distance assignment are used during the population iterations. Specifically, elite selection helps to maintain the best individuals. Fast non-dominated sorting method is used to quickly find superior individuals in populations. Crowding distance assignment improves the search capability of the algorithm. 

\begin{algorithm}[ht]
\caption{\textbf{1:} \,Problem decomposition based genetic algorithm (PDGA)}
\label{algo1}
\begin{algorithmic}[1]
\Require iteration number $G$, population size $Size$, sub-task sequence $Sequence$, simulated binary crossover factor ${eta}_c$, crossover probability $c$, polynomial mutation factor ${eta}_m$, mutation probability $m$, search limit $Limit$.
\Ensure optimal service compositions and service usage schemes $Solut\!\!\:\!\!\:\,ions$.
\State $Pops[\,] \Leftarrow \emptyset $\,;\,/* $Pops[\,]$ is the sequence of populations.*/
\State $Solut\!\!\:\!\!\:\,ions[\,] \Leftarrow \emptyset $\,;\,/* $Solut\!\!\:\!\!\:\,ions[\,]$ is the set of solutions.*/
\State $index \Leftarrow 0$\,;\,/* $index$ is used to guide the populations iteration.*/
\For{each sub-task ${ST}_{\!i}$ in $Sequence$}
    \State $Pops[i] \Leftarrow$ Generate the initial population with $Size$ 
    \Statex \quad\,\,\,\,for ${ST}_{\!i}$\,;\,/* Randomize initial populations.*/
\EndFor
\For{($g=0$\,;\,$g<G$\,;\,$g++$)}
    \State $Solut\!\!\:\!\!\:\,ions[g]$ and $index\Leftarrow$ \textbf{Complete solution gene-}
    \Statex \quad\,\, \textbf {ration($Pops[\,],\,$ $Limit$)}\,;\,/* This function is Algorithm 
    \Statex \quad\,\, \hyperref[algo3]{3} presented in Section \ref{subsec4-4}.*/
    \State $Pops[\,]\Leftarrow$ \textbf{Iteration($Pops[\,],\,index,\,{eta}_c,\, {eta}_m,\,{pr}_c,$}
    \Statex \quad\,\,\,${pr}_m$\textbf{)}\,;\,/* This function is Algorithm \:\hyperref[algo4]{4} presented in
    \Statex \quad\,\,\,Section \ref{subsec4-5}.*/
\EndFor
\State $Solut\!\!\:\!\!\:\,ions[\,] \Leftarrow$ \textbf{Optimal solution selection}\textbf{($Solut\!\!\:\!\!\:\,ions[\,]$)}\,;\,
\Statex/* This function is Algorithm \:\hyperref[algo5]{5} presented in Section \ref{subsec4-6}.*/
\State Output $Solut\!\!\:\!\!\:\,ions[\,]$.
\end{algorithmic}
\end{algorithm}

\subsection{Stratified sequencing approach and search bootstrap function}\label{subsec4-3}

Usually, it is necessary to solicit the opinions of users and the advice of experts before searching for the optimal cloud service composition. Based on the results of the solicitation, multiple service indexes are combined by weighted sum approaches to obtain the final evaluation index: quality of service (QoS). However, it is not easy to determine the weights of each index in the task \cite{2022Thakur(12)}. Opinions and advice are highly dependent on personal experience, and the solicitation process takes additional time and cost. Thus, the stratified sequencing approach is used to prioritize the following three objectives (as previously presented in Section \ref{subsec3-3}): 1) the primary objective is the minimum total completion time; 2) the secondary objective is the lowest total production cost; and 3) the final objective is the minimum number of selected services. The efficiency of the solution search is also improved by the stratified sequencing approach. 

However, the stratified sequencing approach may lead the PDGA to focus too much on reducing total completion time at the expense of other objectives. This does not necessarily correspond to actual demands. Thus, a search bootstrap function is designed in the PDGA to obtain more diverse solutions.
The specific pseudo-code of the function is shown in Algorithm \:\hyperref[algo2]{2}. A search limit, denoted by $Limit$, is set by the user requirements for the objective of total completion time. This limit is used in the search bootstrap function. As the total completion time of a solution approaches this limit, Algorithm \:\hyperref[algo1]{1} will focus on reducing the total production cost and the number of selected services.

\begin{algorithm}[ht]
\caption{\textbf{2:} \,Search bootstrap function}\label{algo2}
\begin{algorithmic}[1]
\Require population $Pop$, search limit $Limit$.
\Ensure individual $Individual$, theoretical minimum completion time $time$.
\State $Individual \Leftarrow \:$null, $time \Leftarrow \:$null\,;\,/* Initialization.*/
\State $time$$[\,] \Leftarrow \emptyset $\,;\,/* $time$$[\,]$ is the sequence of theoretical minimum completion times of individuals in $Pop$.*/
\For{($j=0$\,;\,$j<$ length($Pop$)\,;\,$j++$)}
    \State $t\!\!\:\!\!\:\,\,ime[j]\Leftarrow$ calculate the theoretical minimum comple-
    \Statex \quad\,\,\,tion time of individuals in  $Pop[j]$ by Eq.(\ref{equation:Equation 3})\,;
\EndFor
\State $Mintime\Leftarrow$ min($time[\,]$)\,;
\If{$Mintime\ge Limit$}
    \State $Individual\Leftarrow Pop[time.$index$(Mintime)]$\,;\,/*The ind-
    \Statex \quad\,\, ex of $Mintime$ in $time$$[\,]$ is $time.$index$(Mintime)$.*/
    \State $time \Leftarrow Mintime$\,;
\Else
    \State $Mintime\Leftarrow\infty$\,;
    \For{$time$ in $time[\,]$}
        \If{$time\ge Limit$ \& $time\leq Mintime$}
            \State $Mintime\Leftarrow time$\,;
        \EndIf
    \EndFor
    \If{$Mintime\neq\infty$}            
        \State$Individual\Leftarrow Pop[time.$index$(Mintime)]$\,;
        \State $time \Leftarrow Mintime$\,;
    \Else
        \State$Maxtime\Leftarrow max(time[\,])$\,;
        \State$Individual\Leftarrow Pop[time.$index$(Maxtime)]$\,;
        \State $time \Leftarrow Maxtime$\,;
    \EndIf
\EndIf
\State Output $Individual$ and $time$.
\end{algorithmic}
\end{algorithm}

\subsection{Complete solution generation}\label{subsec4-4}

\hspace{1em}\textbf{Definition 2} \cite{2022Tirkolaee(39)} (Non-dominated solution) Let $X_1$ and $X_2$ be two solutions in the current solution set. Solution $X_2$ is dominated by solution $X_1$ if all objective values of $X_1$ are not worse than the corresponding objective values of $X_2$, and one or more objectives of $X_1$ are better than the corresponding objective values of $X_2$. Solution $X_1$ is a non-dominated solution if $X_1$ is not dominated by any other solution in the current solution set.

Note that a non-dominated solution does not necessarily perform better in all objectives than any other solutions. It only means that there is no other solution that performs better in all objectives than the non-dominated solution $X_1$.
For example, there is a solution set $\{X_1$, $X_2$, $X_3$\}. For two objectives (e.g., the cost and the number of services required), $X_1=(3,2)$, $X_2=(3,3)$, and $X_3=(2,3)$. The solution $X_2$ is dominated by either of solutions $X_1$ and $X_3$. The solution $X_1$ (resp., $X_3$) is not dominated by solution $X_3$ (resp., $X_1$). As a consequence, solutions $X_1$ and $X_3$ are non-dominated solutions in the solution set. 

For each sub-task, a population is created to search for the optimal partial solutions so as to generate complete solutions. The process is as follows. 
First, in each population, select an individual using the search bootstrap function (i.e. Algorithm \:\hyperref[algo2]{2}). Second, among these individuals, select the one with the longest theoretical minimum completion time, denoted by $X_i$, $i\in\{1,\ 2,\:\cdots,\ I\}$. That is, $i$ indicates that $X_i$ is from the $i$th population for ${ST}_{\!i}$. The theoretical minimum completion time of $X_i$ is denoted by $MaxT$. Third, in each population except the $i$th population, calculate the dominance relationship between any two individuals by the fast non-dominated sorting method using Eqs.\:(\ref{equation:Equation 3}-\ref{equation:Equation 5}). Fourth, search for a particular individual having the lowest cost, the fewest number of selected services, and the theoretical minimum completion time less than $MaxT$. Finally, combine these individuals with $X_i$ to form a complete solution, and calculate the total completion time, the total production cost, and the number of selected services of the complete solution by Eqs.\:(\ref{equation:Equation 6}-\ref{equation:Equation 8}).

The specific pseudo-code of the complete solution generation is shown in Algorithm \:\hyperref[algo3]{3}, where the variable $index$ is used to guide the population iteration operation. 
Specifically, the value of $index$ is equal to $i$, which indicates that the individual $X_i$ with the theoretical minimum completion time $MaxT$ is from the $i$th population for ${ST}_{\!i}$. 

\begin{algorithm}[hb]
\caption{\textbf{3:} \,Complete solution generation}\label{algo3}
\begin{algorithmic}[1]
\Require population sequence $Pops[\,]$, search limit $Limit$.
\Ensure complete solution $solut\!\!\:\!\!\:\,ion$, index $index$.
\State $solut\!\!\:\!\!\:\,ion \Leftarrow \emptyset $, $index \Leftarrow$ null, $individual[\,] \Leftarrow \emptyset $, $time[\,] \Leftarrow \emptyset $, $j\Leftarrow 0 $\,;\,/* Initialization.*/
\For{$Pop$ in $Pops[\,]$}
    \State $individual[j]$ and $time[j]\Leftarrow$
    \textbf{Search bootstrap func-}
    \Statex \quad\, \textbf{tion($Pop, Limit$)}\,;\,/* This function is Algorithm \:\hyperref[algo2]{2}
    \Statex \quad\, presented in Section \ref{subsec4-3}.*/
    \If{$time[j]>MaxT$}
        \State $MaxT\Leftarrow time[j]$\,;
        \State $index\Leftarrow j$\,;
    \EndIf
    \State $j++$\,;
\EndFor
\For{($m=0$\,;\,$m<j$\,;\,$m++$)}
    \If{$m\neq index$}
        \State $SmallPop\Leftarrow$ Select the individuals with the theo-
        \Statex \qquad\quad retical minimum completion time (calculated by
        \Statex \qquad\quad Eq.\:(\ref{equation:Equation 3})) less than $MaxT$ in $Pops[m]$\,;
        \State $Individual[m]\Leftarrow$ Select the individual with the
        \Statex \qquad\quad lowest cost and fewer number of selected services
        \Statex \qquad\quad in $SmallPop$\,;
    \EndIf
\EndFor
\State $solut\!\!\:\!\!\:\,ion\Leftarrow Individual[\,]$\,;
\State Output $solut\!\!\:\!\!\:\,ion$ and $index$.
\end{algorithmic}
\end{algorithm}

\subsection{Iteration}\label{subsec4-5}

After generating a complete solution, all populations are iterated once. 
Simulated binary crossover and polynomial mutation are used in each iteration. We assume that there are two arbitrary paternal individuals whose respective genes at a particular position are $x_1$ and $x_2$. After the simulated binary crossover, the offspring genes are $c_1$ and $c_2$ as follows:
\begin{flalign}
\begin{split}
\begin{cases}
c_1=0.5\ast(x_1+x_2)-0.5\ast\beta\ast(x_1-x_2),\\
c_2=0.5\ast(x_1+x_2)+0.5\ast\beta\ast(x_1-x_2),\\
\end{cases}
\end{split}&
\label{equation:Equation 9}
\end{flalign}
where $\beta$ is the spread factor that represents a ratio difference between parent and offspring individuals. The factor $\beta$ is calculated as follows through a random value $r$ in the range [0,1].
\begin{flalign}
\begin{split}
\beta=
\begin{cases}
(2r)^{\tfrac{1}{1+\eta_c}},&\textrm{if}\ r\leq0.5,\\
\left(\dfrac{1}{2-2r}\right)^{\tfrac{1}{1+\eta_c}},&\textrm{if}\ r>0.5,
\end{cases}
\label{equation:Equation 10}
\end{split}&
\end{flalign}
where $\eta_c$ is the spread factor distribution index as a custom parameter. The larger the value of $\eta_c$, the more similar the offspring individuals are to the parent individuals.

For a gene $x$, the gene $m=x+\delta\ast(u-l)$ is obtained by the polynomial mutation of $x$, where $u$ is the upper bound of $x$, $l$ is the lower bound of $x$, and $\delta$ is the perturbation factor calculated as follows through a random value $r$ in the range [0,1].
\begin{flalign}
\begin{split}
\delta=
\begin{cases}
{\left[2r+(1-2r){(1-\delta_1)}^{1+\eta_m}\right]}^{\tfrac{1}{1+\eta_m}-1},\quad\quad\;\;\,\,\, \textrm{if}\ r\leq0.5,\\
1-{\left[2-2r+(2r-1){(1-\delta_2)}^{1+\eta_m}\right]}^{\tfrac{1}{(1+\eta_m)}},\textrm{if}\ r>0.5,
\end{cases}
\vspace{-1em}
\label{equation:Equation 11}
\end{split}&
\end{flalign}
where $\delta_1=(x-l)/(u-l)$, $\delta_2=(u-x)/(u-l)$, and $\eta_m$ is the perturbation factor distribution index as a custom parameter. The larger the value of $\eta_m$, the greater the degree of mutation.

Suppose there are $I$ populations, each with size $Size$. The specific iteration process is as follows. First, cross-mutate each of the $I$ populations by Eqs.\:(\ref{equation:Equation 9}-\ref{equation:Equation 11}) to obtain $I$ offspring populations of size $Size$. Second, according to elite selection, combine each offspring population with its corresponding parent to constitute $I$ current populations of size $2\ast Size$. Third, calculate fitness values for individuals in the current populations using Eqs.\:(\ref{equation:Equation 3}-\ref{equation:Equation 5}). For the individuals in the $i$th population (where $i$ equals to $index$ output by Algorithm \:\hyperref[algo3]{3}), the fitness includes the theoretical minimum completion time calculated by Eq.\:(\ref{equation:Equation 3}), the cost calculated by Eq.\:(\ref{equation:Equation 4}), and the number of selected services calculated by Eq.\:(\ref{equation:Equation 5}). Fitness for other populations includes the cost calculated by Eq.\:(\ref{equation:Equation 4}) and the number of selected services calculated by Eq.\:(\ref{equation:Equation 5}). Finally, for each current population, use the fast non-dominated sorting method with crowding distance assignment to select individuals so as to obtain a new population of size $Size$. The specific pseudo-code of the above iteration is shown in Algorithm \:\hyperref[algo4]{4}.

\subsection{Optimal complete solution selection}\label{subsec4-6}
Instead of updating the maintenance of the complete solution set during the iteration of approach execution, we choose to filter it at the end, in order to reduce the computational cost.
When the maximum number of iterations is reached, a fast non-dominated sorting is performed on all complete solutions to select and output optimal solutions (specifically, non-dominated solutions). After that, the PDGA is finished. The specific pseudo-code of the optimal solution selection is shown in Algorithm \:\hyperref[algo5]{5}.

\begin{algorithm}[ht]
\caption{\textbf{4:} \,Iteration}\label{algo4}
\begin{algorithmic}[1]
\Require population sequence $Pops[\,]$, population size $Size$, index $index$, simulated binary crossover factor ${eta}_c$, crossover probability $c$, polynomial mutation factor ${eta}_m$, mutation probability $m$.
\Ensure population sequence $Pops[\,]$.
\State $j\Leftarrow 0 $\,;\, /* Initialization.*/
\For{$Pop$ in $Pops[\,]$}
    \State $OsPop\Leftarrow$ Generate the offspring population of $Pop$ 
    \Statex \quad\,\,\, by Eqs.\:(\ref{equation:Equation 9}-\ref{equation:Equation 11}) using ${eta}_c$, ${pr}_c$, ${eta}_m$, ${pr}_m$, and $Size$\,;
    \State $Pop\Leftarrow Pop\cup OsPop$\,;
    \If{$j\neq index$}
        \State $Frank[\,]\Leftarrow$ Fast-non-dominated-sort($Pop$) by 
        \Statex \quad\quad\quad$Cost$ and $Num$\,;\,/*$Cost$ is the cost of the individual
        \Statex \quad\quad\quad and $Num$ is the number of selected services in the
        \Statex \quad\quad\quad individual.*/
    \Else
        \State $Frank[\,]\Leftarrow$ Fast-non-dominated-sort($Pop$) by 
        \Statex \quad\quad\quad$Time, Cost,$ and $Num$\,;\,/*$Time$, $Cost,$ and $Num$\ are 
        \Statex \quad\quad\quad calculated by Eqs.\:(\ref{equation:Equation 3}-\ref{equation:Equation 5}).*/
    \EndIf
    \State $NewPop\Leftarrow \emptyset $ and $k\Leftarrow 0$\,;
    \While{($|NewPop|+|Frank[k]|\leq Size$)}
        \State $NewPop\Leftarrow NewPop\cup Frank[k]$\,;\,/*Retain the go-
        \Statex \quad\quad\quad od individuals.*/
        \State $k++$\,;
    \EndWhile
    \State $Distance\Leftarrow$ Crowding-distance-assignment
    \Statex \quad\,\,\,($Frank[k]$)\,;
    \State Sort($Frank[k]$) by $Distance$\,;\,/* Descending order*/
    \State $NewPop\Leftarrow NewPop\cup Frank[k][0:(Size-$
    \Statex \quad\,\,\,$|NewPop|)]$\,;\,/*Retain the first ($Size-|NewPop|$)
    \Statex \quad\,\,\,individuals in $Frank[k]$.*/
    \State $Pops[j]\Leftarrow NewPop$\,;
    \State $j++$\,;
\EndFor
\State Output $Pops[\,]$.
\end{algorithmic}
\end{algorithm}

\vspace{-1em}
\begin{algorithm}[ht]
\caption{\textbf{5:} \,Optimal solution selection}\label{algo5}
\begin{algorithmic}[1]
\Require the set of complete solutions $Solut\!\!\:\!\!\:\,ions[\,]$.
\Ensure optimal solutions $O\!\!\:\!\!\:\,ptimalSolut\!\!\:\!\!\:\,ions[\,]$.
\State $O\!\!\:\!\!\:\,ptimalSolut\!\!\:\!\!\:\,ions[\,]\Leftarrow \emptyset $\,; \,/* Initialization.*/
\State $Frank[\,]\Leftarrow$ Fast-non-dominated-sort($Solut\!\!\:\!\!\:\,ions[\,]$) by
\Statex ${Time}_{total}$, ${Cost}_{total}$, and ${Num}_{total}$\,;\,/*${Time}_{total}$, ${Cost}_{total}$,
\Statex and ${Num}_{total}$ are calculated by Eqs.\:(\ref{equation:Equation 6}-\ref{equation:Equation 8}).*/
\State $O\!\!\:\!\!\:\,ptimalSolut\!\!\:\!\!\:\,ions[\,]\Leftarrow Frank[0]$\,;
\State Output $O\!\!\:\!\!\:\,ptimalSolut\!\!\:\!\!\:\,ions[\,]$.
\end{algorithmic}
\end{algorithm}
\vspace{-1em}

\section{Case study}\label{sec5}

\begin{figure}[ht]
\centering 
\includegraphics[scale=0.6]{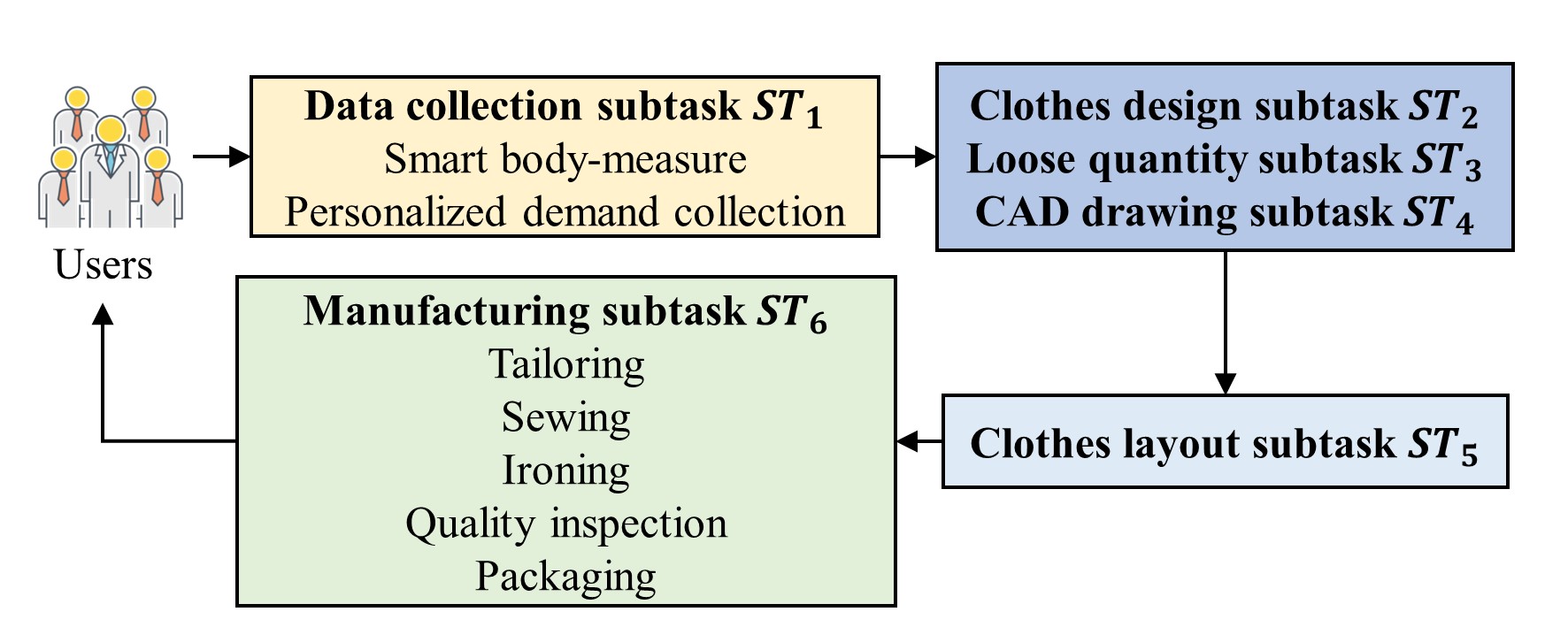}
\caption{\small Process of the CMfg clothing customization production}
\label{fig:Fig.5}
\end{figure}

\begin{table}[h]
\caption{\small The expected output of each sub-task}\label{tab1}%
\tabcolsep=0.1cm
\renewcommand\arraystretch{1.2}
\begin{tabular*}{\hsize}{@{}@{\extracolsep{\fill}}cccc@{}}
\Xhline{0.8pt}
Sub-task&${ST}_1$&${ST}_2$&${ST}_3$\\
\hline
Output&\makecell[c]{Body data and\\customization needs}&\makecell[c]{Clothes\\design data}&\makecell[c]{Clothes design\\data with (a)}\\
\Xhline{0.8pt}
Sub-task&${ST}_4$&${ST}_5$&${ST}_6$\\
\hline
Output&\makecell[c]{Clothes design\\data with (a) and (b)}&\makecell[c]{Clothes design data\\with (a), (b), and (c)}&Clothing\\
\Xhline{0.8pt}
\end{tabular*}
\footnotetext{(a) loose quantity data
(b) CAD drawing data
(c) clothes layout data}
\end{table}

In this section, we apply the PDGA to a CMfg clothing customization production system which is modified from \cite{2022WangMin(2)} and whose entire process is shown in Fig.\:\ref{fig:Fig.5}. 
First, the user data is collected by the data collection sub-task ${ST}_1$, which includes the body measurements and the personalized demands. Second, the process enters the automated design phase, including the clothes design sub-task ${ST}_2$, the loose quantity sub-task ${ST}_3$, and the CAD drawing sub-task ${ST}_4$. Then, the process proceeds to the clothes layout sub-task ${ST}_5$, followed by the manufacturing sub-task ${ST}_6$ which includes tailoring, sewing, ironing, quality inspection, and packaging. Table\:\ref{tab1} shows the expected output for each sub-task.

\begin{table}[h]
\caption{\small Parameters of experimentation}\label{tab2}%
\renewcommand\arraystretch{1.5}
\begin{tabular*}{\hsize}{@{}@{\extracolsep{\fill}}c|ccc|cc|cc@{}}
\Xhline{0.8pt}
Sub-task&\multicolumn{3}{c|}{${ST}_1$}&\multicolumn{2}{c|}{${ST}_2$}&\multicolumn{2}{c}{${ST}_3$}\\
\hline
\makecell[c]{Candidate\\service}&${CS}_{1,1}$&${CS}_{1,2}$&${CS}_{1,3}$&${CS}_{2,1}$&${CS}_{2,2}$&${CS}_{3,1}$&${CS}_{3,3}$\\
\hline
Time&1&0.8&1.1&10&15&3&2\\
\hline
Cost&1&1.2&0.9&12&10&1&1.5\\
\Xhline{0.8pt}
Sub-task&\multicolumn{3}{c|}{${ST}_3$}&\multicolumn{2}{c|}{${ST}_5$}&\multicolumn{2}{c}{${ST}_6$}\\
\hline
\makecell[c]{Candidate\\service}&${CS}_{4,1}$&${CS}_{4,2}$&${CS}_{4,3}$&${CS}_{5,1}$&${CS}_{5,2}$&${CS}_{6,1}$&${CS}_{6,2}$\\
\hline
Time&25&28&22&6&8&50&45\\
\hline
Cost&2&1.6&2.4&2&1.6&15&18\\
\Xhline{0.8pt}
\end{tabular*}
\end{table}

Table\:\ref{tab2} describes the candidate services for each sub-task. It also provides the time and cost metrics for a single use of each candidate service.

\begin{figure}[ht]
\centering 
\includegraphics[scale=0.58]{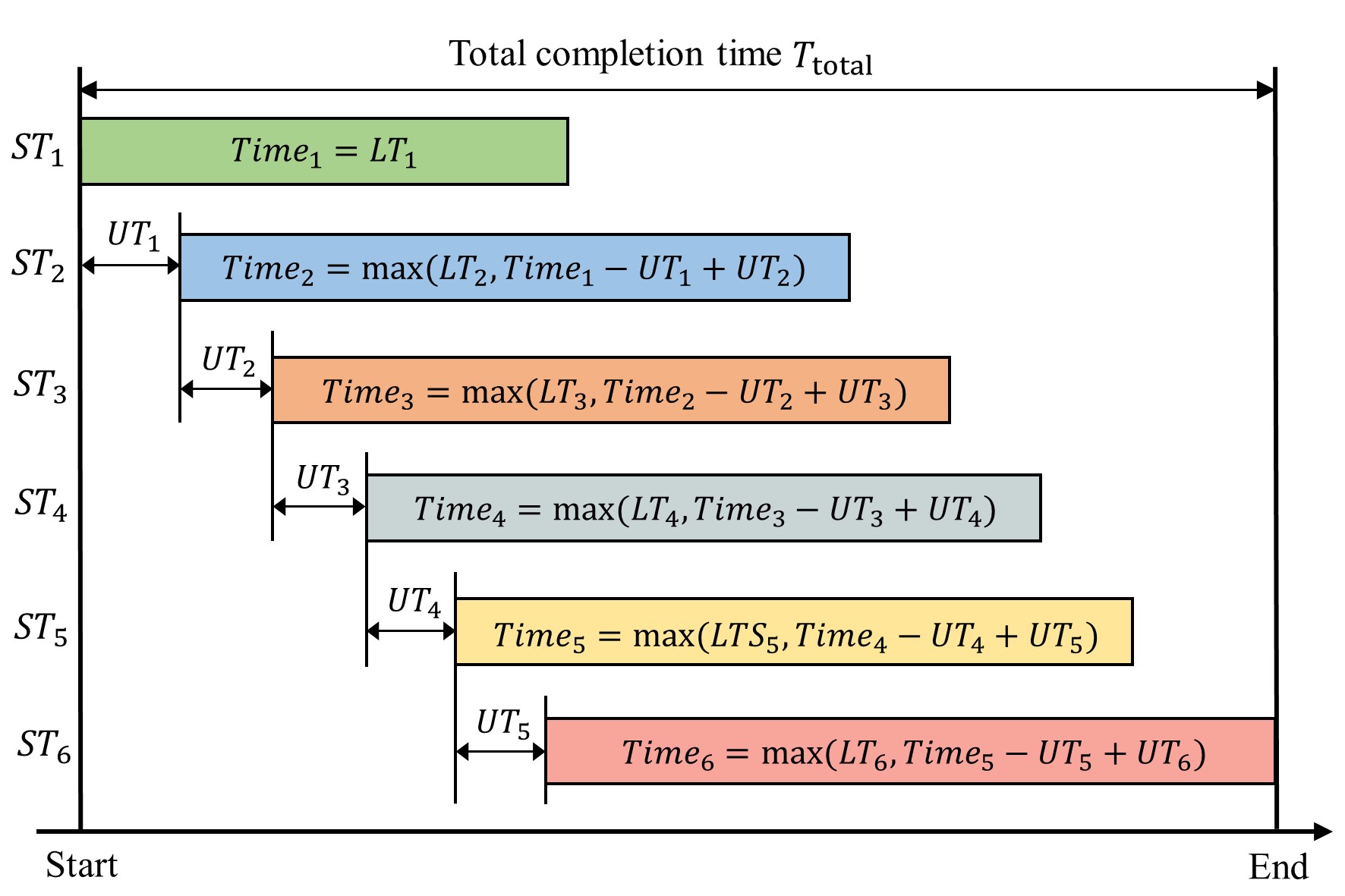}
\caption{\small Gantt chart of the CMfg clothing customization production}
\label{fig:Fig.6}
\end{figure}

As shown in Fig.\:\ref{fig:Fig.6}, for the clothing customization example, the completion time {${Time}_1$} of the first sub-task ${ST}_1$ is equal to the cumulative usage time ${LT}_1$ of ${LCS}_1$. For the other sub-tasks ${ST}_{\!i}$, the completion time {${Time}_i$} is equal to $\max({LT}_{i},\ {Time}_{i-1}-{UT}_{i-1}+{UT}_{i})$, $i=2,\ 3,\:\cdots,\ 6$.
Using Eq.\:(\ref{equation:Equation 6}), the total completion time ${Time}_{total}={Time}_6+\sum_{i=1}^{5}{UT}_i$.

To verify the effectiveness of the method developed in the present work, we conduct experiments using the PDGA presented in Section \ref{sec4} and the multi-objective optimization genetic algorithm NSGA-II. The number of iterations is set to 100 (resp., 200) for the PDGA (resp., NSGA-II). Both NSGA-II and PDGA include elite selection. Therefore, the crossover and mutation probabilities are set to 100\% for a better search efficiency. The spread factor distribution index $\eta_c$ (resp., perturbation factor distribution index $\eta_m$) is set to 0.1 (resp., 0.01). The product quantity in the order is set to 1000. Some specific settings for the experiments are shown in Table\:\ref{tab3}. Both algorithms are implemented in Python3 and tested on a computer with a 2.20 GHz Intel Core i7 8750H CPU and 8.0 GB of RAM running 64-bit Windows 10 operating system.

\begin{table}[h]
\caption{\small Some specific settings for the experiments}\label{tab3}%
\tabcolsep=0.1cm
\renewcommand\arraystretch{1.2}
\begin{tabular*}{\hsize}{@{}@{\extracolsep{\fill}}cccccccc@{}}
\Xhline{0.8pt}
Execution&1&2&3&4&5&6&7\\
\hline
Algorithm&NSGA-II&PDGA&PDGA&PDGA&PDGA&PDGA&PDGA\\
\hline
\makecell[c]{Iteration\\number}&200&100&100&100&100&100&100\\
\hline
$Limit$&-&0&24000&26000&28000&30000&32000\\
\Xhline{0.8pt}
Execution&8&9&10&11&12&13&14\\
\hline
Algorithm&PDGA&PDGA&PDGA&PDGA&PDGA&PDGA&PDGA\\
\hline
\makecell[c]{Iteration\\number}&100&100&100&100&100&100&100\\
\hline
$Limit$&34000&36000&38000&40000&42000&44000&46000\\
\Xhline{0.8pt}
\end{tabular*}
\end{table}

\begin{figure*}[hb]
\begin{minipage}[t]{0.315\linewidth}
\caption{Total completion time and total production cost of solutions}
\label{fig:Fig.7}
\end{minipage}
\hfill
\begin{minipage}[t]{0.7\linewidth}
\flushright
\vspace{-1em}
\includegraphics[scale=0.6]{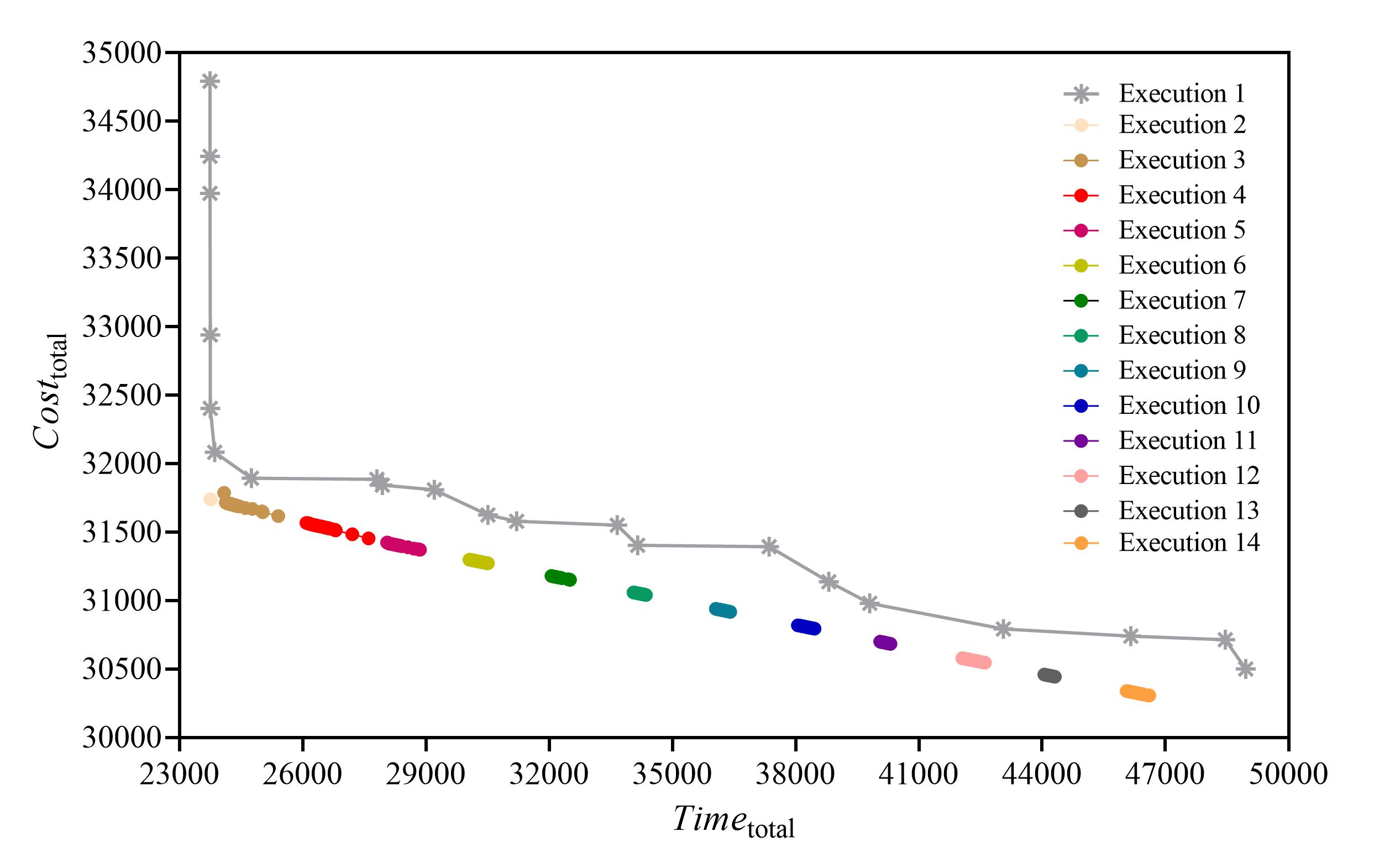}
\end{minipage}
\end{figure*}
\FloatBarrier

Each execution of the experiment produces a set of non-dominated solutions. Fig.\:\ref{fig:Fig.7} where each point represents a solution, shows the total completion time and the total production cost of the solutions obtained by each execution. Table\:\ref{tab4} shows the number of solutions obtained by each execution. It is easy to see that the solutions produced by the PDGA are denser compared with those by NSGA-II, due to the stratified sequencing approach strategy used in the PDGA. The strategy allows the PDGA to better search for optimal solutions. Clearly, the solutions produced by the PDGA are better in both aspects of total completion time and total production cost than those by the NSGA-II.

\begin{table}[h]
\caption{\small The number of solutions in each execution}\label{tab4}%
\tabcolsep=0.1cm
\renewcommand\arraystretch{1.2}
\begin{tabular*}{\hsize}{@{}@{\extracolsep{\fill}}cccccccc@{}}
\Xhline{0.8pt}
Execution&1&2&3&4&5&6&7\\
\hline
Algorithm&NSGA-II&PDGA&PDGA&PDGA&PDGA&PDGA&PDGA\\
\hline
\makecell[c]{Number of\\solutions}&21&3&20&23&13&10&8\\
\Xhline{0.8pt}
Execution&8&9&10&11&12&13&14\\
\hline
Algorithm&PDGA&PDGA&PDGA&PDGA&PDGA&PDGA&PDGA\\
\hline
\makecell[c]{Number of\\solutions}&7&8&9&6&10&6&11\\
\Xhline{0.8pt}
\end{tabular*}
\end{table}

\begin{figure*}[ht]
\begin{minipage}[t]{0.3\linewidth}
\caption{Average number of selected services}
\label{fig:Fig.8}
\end{minipage}
\hfill
\begin{minipage}[t]{0.7\linewidth}
\flushright
\vspace{-1em}
\includegraphics[scale=0.6]{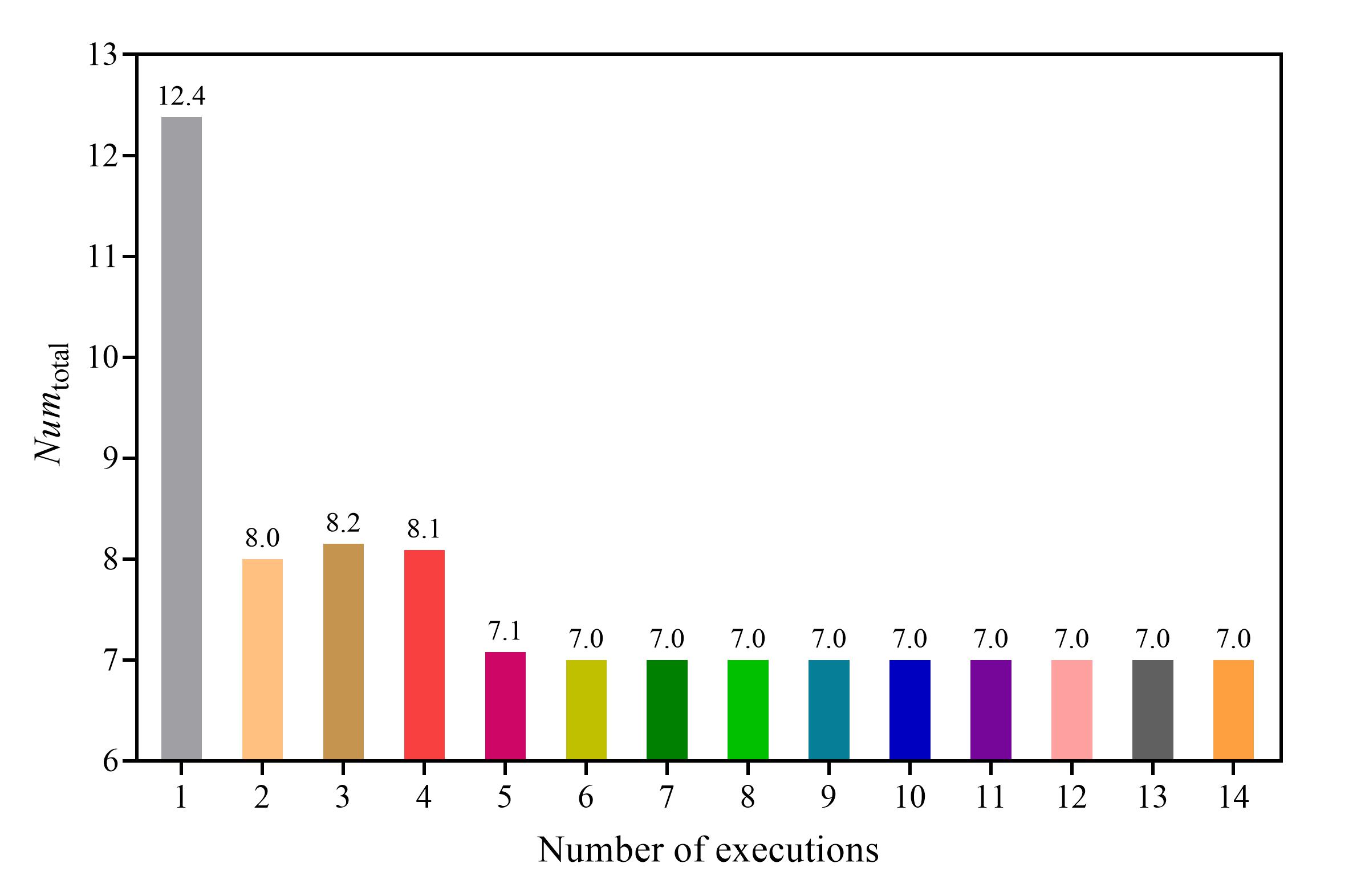}
\end{minipage}
\end{figure*}

Fig.\:\ref{fig:Fig.8} shows the average number of selected services in the solutions obtained by each execution. The algorithm employed in the first execution is the NSGA-II, which is used for comparative purposes. The others are the PDGA with different values of $Limit$. Clearly, the solutions obtained by the PDGA are with fewer selected services than those obtained by the NSGA-II. The results indicate that the solutions with respect to the PDGA need fewer service resources compared with those of NSGA-II. This empirical evidence points to the efficiency of our method.

\begin{figure*}[ht]
\begin{minipage}[t]{0.3\linewidth}
\caption{Time consumption of each execution}
\label{fig:Fig.9}
\end{minipage}
\hfill
\begin{minipage}[t]{0.7\linewidth}
\flushright
\vspace{-1em}
\includegraphics[scale=0.6]{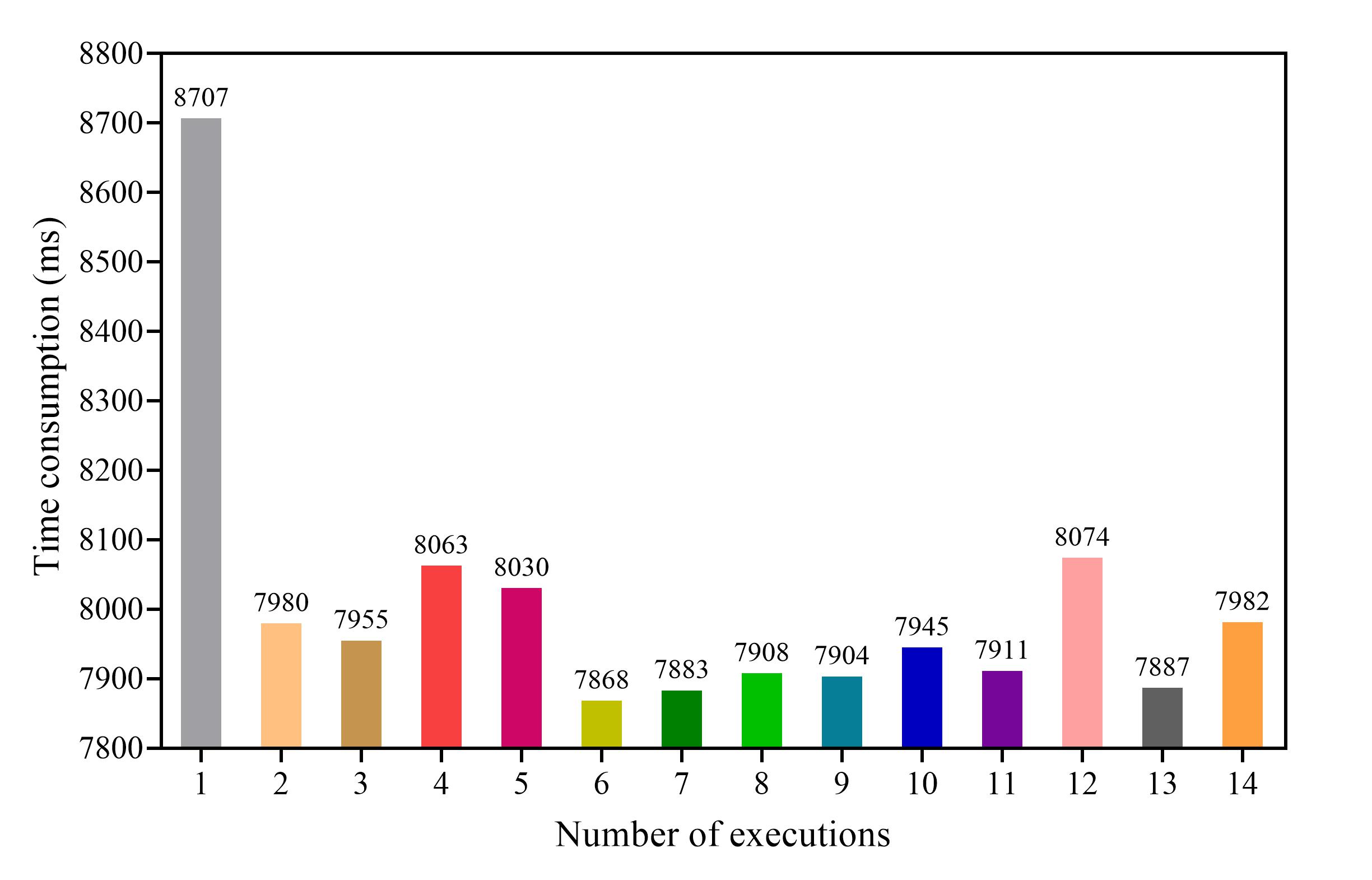}
\end{minipage}
\end{figure*}

Fig.\:\ref{fig:Fig.9} shows the time consumption of each execution. The first execution employs the NSGA-II, which is again used for comparative purposes. The other executions are of the PDGA with different values of $Limit$. To ensure the data accuracy, we calculated the average computation time. The results indicate that the PDGA finds better solutions in less time than the NSGA-II.

In summary, the PDGA produces better solutions in less time compared with the NSGA-II. Additionally, the PDGA can be configured with different $Limit$ values based on completion time requirements, resulting in solutions that address user demands more effectively.

\FloatBarrier

\FloatBarrier

\section{Comparison}\label{sec6}

This section aims to compare the difference between our work and some representative studies for optimal CMfg service composition selection. The mass comparisons are shown in Table\:\ref{tab5}.

\vspace{-1em}
\begin{table}[h]
\caption{\small Comparison of methods of handling the SCOS}\label{tab5}%
\begin{tabular*}{\hsize}{@{}@{\extracolsep{\fill}}lcccccc@{}}
\toprule
Work&(a)&(b)&(c)&(d)&(e)\\
\midrule
\cite{2017ZhouJiajun(23)}&\begin{tabular}[c]{@{}c@{}}\thead{Ordinary\\production}\end{tabular}&\thead{Single\\service}&\begin{tabular}[c]{@{}c@{}}(f)\end{tabular}&Yes&No\\

\cite{2018Que(28)}&\begin{tabular}[c]{@{}c@{}}\thead{Ordinary\\production}\end{tabular}&\thead{Single\\service}&\begin{tabular}[c]{@{}c@{}}(f)\end{tabular}&Yes&No\\

\cite{2020ding(11)}&\begin{tabular}[c]{@{}c@{}}\thead{Ordinary\\production}\end{tabular}&\thead{Single\\service}&\begin{tabular}[c]{@{}c@{}}(f)\end{tabular}&Yes&No\\

\cite{2022WangMin(2)}&\begin{tabular}[c]{@{}c@{}}\thead{Customized\\production}\end{tabular}&\thead{Multiple\\services}&(g)&No&No\\

Ours&\begin{tabular}[c]{@{}c@{}}\thead{Customized\\production}\end{tabular}&\thead{Multiple\\services}&(g)&No&Yes\\
\botrule
\end{tabular*}
\footnotetext{(a) For what problem?
(b) Single service or multiple services for each sub-task?
(c) What is the type of optimization?
(d) Is additional weight needed?
(e) Does the optimization consider the objective of using fewer service resources?
(f) Converting multi-objective optimization to single-objective optimization
(g) Multi-objective optimization}
\end{table}

In particular, the present work is different from \cite{2017ZhouJiajun(23)}, \cite{2018Que(28)}, and \cite{2020ding(11)} since there is no inclusion between the problem classes studied by them and ours. Compared with the previous work of Wang et al. \cite{2022WangMin(2)}, the present work considers not only the collaboration among services with the same functionality, but also the amount of service resources. Additionally, the method developed in the present work can quickly find the optimal service compositions and usage schemes. In contrast, the optimization of service composition is not addressed in the reference \cite{2022WangMin(2)}.

\section{Conclusion and future work}\label{sec7}

For the cloud manufacturing (CMfg) service system that handles personalized requirements, this paper proposed a framework with collaboration among similar services to improve the efficiency and reduce the cost of the cloud manufacturing customized production (CMCP). Then, the optimal service composition and service usage scheme of CMCP is presented. Finally, a problem decomposition based genetic algorithm using the stratified sequencing approach was proposed to achieve the optimization of the CMCP. The results show that the solution obtained by this algorithm has the shortest total completion time and lower total production cost, and requires fewer services. 

In the future, the dependency among services and the uncertainty of service delivery in CMfg service systems as well as the multi-order problem will be addressed to study the CMCP. In addition, we plan to develop a method to calculate quickly and accurately the time required for mass customization production, taking into account multiple services with the same functionality working together.
\\
\\
\small\noindent\textbf{Acknowledgement} This work was supported by the National Natural Science Foundation of China [grant number 62072260].
\\
\\
\noindent\textbf{Authorship contribution} Hao Yue contributed to main idea, formal analysis, methodology, literature search, data analysis, figures, and writing—original draft preparation—review \& editing. Yingtao Wu contributed to main idea, data analysis, figures, and writing—review \& editing. Min Wang contributed to funding acquisition, supervision, writing—review \& editing. Hesuan Hu contributed to writing—review \& editing. Weimin Wu contributed to validation, visualization, writing—review \& editing. Jihui Zhang contributed to funding acquisition, resources. Provide the same order of author in both the system and the manuscript file and the meta-data.
\\
\\
\noindent\textbf{Funding} National Natural Science Foundation of China [grant number 62072260]
\\
\\
\noindent\textbf{Data availability} Data and materials are available on request from the authors.

\section*{Declarations}

\noindent\textbf{Ethical approval} Not applicable.
\\
\\
\noindent\textbf{Consent to participate} Not applicable.
\\
\\
\noindent\textbf{Consent to publish} All authors have agreed for authorship, read, and approved the manuscript, and given consent for submission and subsequent publication of the manuscript. The authors guarantee that the contribution to the work has not been previously published elsewhere.
\\
\\
\noindent\textbf{Competing interests} The authors declare no competing interests.

\bibliography{sn-bibliography}

\end{document}